# Operation of a high purity germanium crystal in liquid argon as a Compton suppressed radiation spectrometer

John L. Orrell,[a,][*] Craig E. Aalseth,[a] John F. Amsbaugh,[b] Peter J. Doe,[b]

Todd W. Hossbach[a,1]

[a]*Pacific Northwest National Laboratory, 902 Battelle Boulevard, Richalnd, WA 99352, USA*

[b]*Center for Experimental Nuclear Physics and Astrophysics, Univeristy of Washington, Seattle, WA 98105, USA*

**Abstract**

A high purity germanium crystal was operated in liquid argon as a Compton suppressed radiation spectrometer. Spectroscopic quality resolution of less than 1% of the full-width half maximum of full energy deposition peaks was demonstrated. The construction of the small apparatus used to obtain these results is reported. The design concept is to use the liquid argon bath to both cool the germanium crystal to operating temperatures and act as a scintillating veto. The scintillation light from the liquid argon can veto cosmic-rays, external primordial radiation, and gamma radiation that does not fully deposit within the germanium crystal. This technique was investigated for its potential impact on ultra-low background gamma-ray spectroscopy. This work is based on a concept initially developed for future germanium-based neutrinoless double-beta decay experiments.

————
[*] Corresponding author. Tel.: +1-509-376-4361; fax: +1-509-376-8002; e-mail: john.orrell@pnl.gov.
[1] Present address: University of South Carolina, Columbia, SC 29208.

# 1. Introduction

Based on an idea by G. Heusser [1] and an initial design concept by S. Schoenert [2], an experimental apparatus to operate a high purity germanium (HPGe) crystal in a cryogenic liquid argon (LAr) bath as a Compton-suppressed, gamma-ray spectrometer has been constructed. The Compton suppression of HPGe gamma-ray spectra is obtained by monitoring the LAr volume for scintillation light. The technical advantages of a LAr Compton suppression system are:

- Direct cooling of the crystal via the LAr bath,
- Minimal absorbing material between the HPGe and the scintillating LAr, and
- Reduction of potentially radiologically impure HPGe structural support material.

This detector concept may provide advantages to ultra-low-background gamma-ray spectrometers. The sensitivity of ultra-low-background gamma-ray spectrometers is produced by a combination of low-background construction materials, high density passive shielding (e.g. lead), active cosmic-ray veto shielding, and radon suppression systems [3-5]. Locating these detector systems underground provides additional shielding [6], but at the expense of reducing the ease of accessibility. Even with these methods, the gamma-ray spectra may be dominated by the Compton continuum. The Compton continuum is a result of gamma-rays depositing less than their full energy in the active volume of the detector [7]. Thus higher energy gamma-rays produce a Compton continuum obscuring the full energy peaks of lower energy gamma-rays. This is a sample associated background [3] such that even in a perfectly "background free" detector, the Compton continuum produced by radionuclides in the sample can be the limiting factor in identifying and quantifying all radionuclides actually present.

Compton continuum suppression is possible if a secondary gamma-ray detector is placed around the primary gamma-ray spectrometer. The secondary detectors are most often NaI(Tl) or bismuth germinate (BGO) (See references in Ref. [8]). Ultra-low-background gamma-ray spectrometers can benefit from Compton suppression, if no radioactivity is introduced in the construction. Liquid argon can provide a low-background environment and implement Compton suppression via the ~40,000 scintillation photons generated per MeV [9]. Many of these same considerations initiated S. Schoenert's original investigation of this technique for its impact on future germanium-based neutrinoless double-beta decay experiments. This article details the construction and operation of a LAr based Compton suppressed HPGe gamma-ray spectrometer.

# 2. Experimental Apparatus

A 24% intrinsic p-type germanium blind-hole coaxial detector was used in this work. Nearly all HPGe detectors are operated in vacuum at liquid nitrogen temperatures (77º K). For this experiment, a low-mass crystal mount was fabricated using materials replaceable by low-activity counterparts. The LAr bath provides an appropriate crystal operating temperature (87º K) and an inert noble gas environment. In addition to providing Compton suppression, the LAr acts as a shield and active veto against external gamma-ray backgrounds. The crystal is suspended in the center of the LAr volume of 46-cm depth and 23-cm diameter. The crystal operating bias was selected based on the measured leakage current. An Electron Tubes 9357 KFLB 200 mm diameter photomultiplier tube (PMT) was used to detect LAr scintillation light. The PMT's peak sensitivity is to 380 nm wavelength photons when operated in liquid nitrogen.

There were two operational periods. During the first period, the face of the PMT was submerged in the LAr without any additional light collection methods employed. During the second period, a reflective foil (3M Radiant Mirror Film VM2000 [10,11]) lined the walls of the Dewar. This foil is 98% reflective to wavelengths from approximately 425 nm to 1000 nm. Typically a coating, such as tetraphenyl butadiene (TPB) [12], is used to wavelength shift the LAr scintillation light from its 129 nm emission wavelength to wavelengths reflected by the film. Although no wavelength-shifting coating had been applied, the foil showed a "light-piping" effect that increased the PMT's signal response.

### 3. Chronology of Operation and Results

In this analysis, raw counts-per-channel data histograms are fit to determine the peak energy resolution. Fits use the form

$$N = (mx + b) + \left(\frac{A}{\sigma\sqrt{2\pi}}\right)e^{-\frac{(x-x_0)^2}{2\sigma^2}}$$

where $N$ is the number of counts in channel $x$. The fit assumes a linear background continuum (slope $m$ and intercept $b$) plus a Gaussian distributed peak (magnitude $A$, mean $x_0$, and standard deviation $\sigma$). Full-width half-maximum (FWHM) peak resolution values are reported as a percentage of the peak mean

$$\text{FWHM (\%)} = 100 \times \left(2\sqrt{2\ln 2}\right)\left(\sigma/x_0\right).$$

Prior to operation in LAr, the HPGe detector was tested in its original commercial cryostat. Operated at 1500 Volts, the best FWHM peak energy resolution was 0.12% for the 2614.5-keV peak of $^{208}$Tl. During LAr operation, the distance between the crystal and the signal amplifying field effect transistor (FET) is tens of centimeters. Introducing 76 cm of additional wire length into the commercial cryostat, the FWHM peak resolution of the 2614.5-keV peak broadened to 0.14%.

Figure 1 shows an energy-scaled, re-binned histogram from the first operation of the HPGe in LAr at a bias of 1700 Volts. In all cases reported here, the energy scale is estimated from a linear fit to the three prominent gamma-ray peaks of a $^{207}$Bi source. The LAr scintillation light based Compton suppression system was not yet functioning during this test. Without optimizing the system (e.g. noise suppression and energy calculation optimization), resolutions from 0.79% to 0.24% were obtained. From this first operational run it was concluded the crystal and assembly were operating as expected.

Figure 2 shows an energy-scaled, re-binned histogram from the second operation of the HPGe in the LAr at a bias of 1400 Volts. The spectra shown were acquired in two consecutive runs lasting 30 minutes each. The upper spectrum lacks the LAr scintillation veto. The lower spectrum applies a LAr scintillation veto via a hardware veto-gate. These two figures demonstrate first that good resolution can be obtained from an HPGe crystal operated in LAr and second that the LAr based Compton suppression technique works.

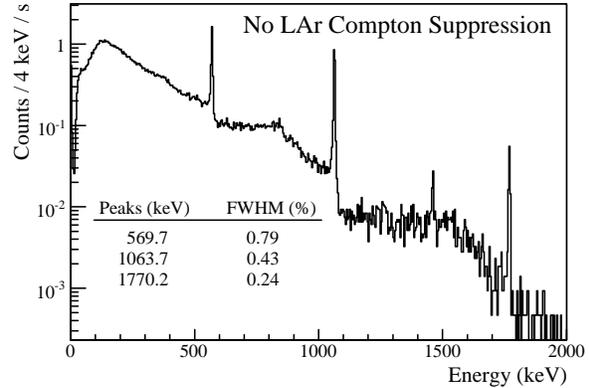

| Peaks (keV) | FWHM (%) |
|---|---|
| 569.7 | 0.79 |
| 1063.7 | 0.43 |
| 1770.2 | 0.24 |

Fig. 1. Demonstration of spectroscopic quality energy resolution from a germanium detector operated in liquid argon.

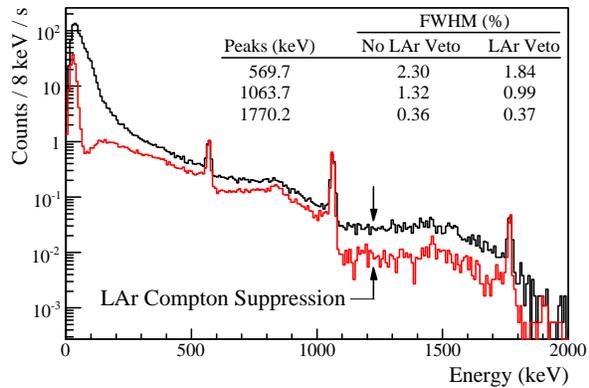

| | FWHM (%) | |
| Peaks (keV) | No LAr Veto | LAr Veto |
|---|---|---|
| 569.7 | 2.30 | 1.84 |
| 1063.7 | 1.32 | 0.99 |
| 1770.2 | 0.36 | 0.37 |

Fig. 2. Demonstration of Compton-continuum suppression in a germanium detector using a scintillating liquid argon bath.

However, the crystal's performance was degraded between the two operational periods. Not only are the FWHM peak resolutions (1.84% to 0.37%) worse than the first operation, the leakage current of the crystal forced operation at reduced bias. Surface contamination resulting in conductive paths is a major culprit in the degradation of HPGe detectors. Two possible causes of this degradation have been identified. The lid of the LAr Dewar does not create a gas seal so a larger containment barrel was used to create a gas buffer region. Filling the Dewar creates boil-off Ar gas in the containment barrel, maintaining an Ar gas barrier between the LAr and atmosphere. In this process, it is possible atmospheric gases condense in the LAr and/or on the HPGe crystal. Another possibility for the degradation of the crystal is from the vacuum bell jar storage used between operation periods. A post-operation inspection revealed the vacuum line to the bell jar was leaking to atmosphere. It is not clear if the degradation of the crystal is due to condensation of contaminates in the LAr bath or to approximately one month of exposure to atmosphere. Future work is directed toward stringent control of the crystal's environment.

## 4. Discussion and Conclusions

Returning to the analysis of the experimental results, the [207]Bi sample used as a gamma-ray source produces three primary paired gamma-ray cascades (in keV): (1770 & 570), (1442 & 897), and (1063 & 750). The pairs of gamma-rays in the latter two low-energy cascades have correlated emission directions [13]. Forward emission correlation can create an enhancement to the LAr veto condition [14]. However, the 1770 keV - 570 keV cascade is isotropic. Additionally, because the source was located 54 cm from the LAr volume, less than 1% solid angle from the [207]Bi intersects the LAr volume. Thus, correspondingly, less than 1% of the 1770 keV - 569 keV gamma-ray pairs will both enter the LAr volume. Thus is it reasonable to simulate the 1770-keV gamma ray as independent from the other [207]Bi gamma rays.

A FLUKA Monte Carlo simulation including only the gross features of the experimental apparatus, namely the LAr and HPGe volumes, was performed. The simulation analysis used energy deposition scoring to understand the relative fractions of gamma-rays that interact in the HPGe and LAr. The Compton suppression factor for the continuum induced by the 1770-keV gamma ray is measured as an integral of counts in the 1200-keV to 1400-keV energy region. Assuming a LAr detection threshold in the simulation of 10 keV deposited energy, the expected Compton suppression factor is 3.8. The measured Compton suppression factor was 3.4. Changing the assumed detection threshold in the simulation to 45 keV reproduces the measured suppression factor. This simulation demonstrates the Compton continuum suppression measured using this experimental apparatus is consistent with expectations based on the volume of LAr present around the HPGe crystal.

## Acknowledgments

We thank S. Schoenert for encouragement and informative discussions on our design concept. We acknowledge our technical staff at Pacific Northwest National Laboratory (PNNL) and University of Washington's Center for Experimental Nuclear Physics and Astrophysics (UW/CENPA). We thank the U.S. Department of Energy's (U.S. DOE) Institute for Nuclear Theory for hosting the summer session on Underground Science (INT-05-2a), during which some of this work was completed. This research was supported by the Laboratory Fellows Initiative at PNNL (U.S. DOE Contract #DE-AC06-76RL01830) and by UW/CENPA (U.S. DOE Grant #DE-FG02-97ER41020/A000).